\documentclass[preprint,amsmath,showpacs]{revtex4}
\begin{document}

\title{Enhanced quantum tunnelling induced by disorder}
\author{Jean Heinrichs}
\email{J.Heinrichs@ulg.ac.be} \affiliation{D\'{e}partement de
Physique, B{5a}, Universit\'{e} de Li\`{e}ge, Sart Tilman, B-{4000}
Li\`{e}ge, Belgium}
\date{\today}

\begin{abstract}
We reconsider the problem of the enhancement of tunnelling of a
quantum particle induced by disorder of a one-dimensional tunnel
barrier of length $L$, using two different approximate analytic
solutions of the invariant imbedding equations of wave propagation
for weak disorder. The two solutions are complementary for the
detailed understanding of important aspects of numerical results on
disorder-enhanced tunnelling obtained recently by Kim et al. (Phys.
rev. B{\bf 77}, 024203 (2008)). In particular, we derive
analytically the scaled wavenumber $(kL)$-threshold where
disorder-enhanced tunnelling of an incident electron first occurs,
as well as the rate of variation of the transmittance in the limit
of vanishing disorder. Both quantities are in good agreement with
the numerical results of Kim et al. Our non-perturbative solution of
the invariant imbedding equations allows us to show that the
disorder enhances both the mean conductance and the mean resistance
of the barrier.
\end{abstract}


\maketitle

Growing attention has been devoted in recent years to an intriguing
phenomenon occurring when a quantum-mechanical particle of energy
$E$ impinges on a disordered one-dimensional tunnel barrier of mean
height $V>E$.  Indeed it was found that the transmission coefficient
of the barrier is increased by the effect of a weak
disorder\cite{1,2,3}.  This is surprising since one would expect
that the additional scattering of an incoming electron by the
disorder (e.g. random impurity potentials) would reduce the
transmission rather than enhancing it.

The interest in the study of the transmission of quantum particles
through disordered one-dimensional barriers\cite{1,2,3} was
triggered by an earlier study of transmission of scalar waves
through disordered one-dimensional dielectric media\cite{4}.  In
this case, the increase of the transmission coefficient with
increasing disorder was first observed for frequencies of the
incident wave lying in the gap of the band structure of the periodic
medium in the absence of disorder\cite{4}.  The similarity between
the quantum barrier- and the dielectric medium problems comes from
the fact that wave frequencies in the gap of the dielectric medium
correspond to evanescent wave solutions of the wave equation, which
are analogous to the evanescent states solutions of the
Schr\"{o}dinger equation for energies $E<V$ in the potential barrier
problem.

Prior to the studies of transmission by disordered potential
barriers in Refs. \cite{1,2,3}, the author developed an extensive
analytic study of resistance- and reflectance distributions in
one-dimensional disordered tunnel barriers in the context of
electron localization\cite{5,6}.  These studies, based on the
invariant imbedding method\cite{7,8}, were motivated by the
development of a fully probabilistic scaling theory of
localization\cite{9} in these systems where localization effects
coexist with the familiar barrier penetration (tunnelling)
effects\cite{5,6}.

The purpose of this Brief Report is first to apply the approximate
solution for the reflection amplitude of a disordered barrier
obtained in Ref.\cite{5} for deriving analytic expressions for mean
disorder- enhanced transmission coefficients and to compare the
results with previous works\cite{1,2,3}.  In particular, we are
interested in understanding an important new aspect of the
transmission of a disordered barrier of length $L$ revealed by the
numerical results of Kihong Kim et al.\cite{3}, namely the existence
of a threshold value of the scaled particle wavenumber $kL$ above
which transmission is enhanced by the disorder while being reduced
by the disorder below it.

On the other hand, it is found that the thresholds for enhanced
transmission obtained numerically\cite{3} vary significantly with
the relative barrier hight $V/E$.  In order to understand this
behaviour, it is necessary to develop a solution of the invariant
imbedding equations for arbitrary barriers $V/E>1$, in parallel to
the solution in\cite{5} which is specific to the case $V/E=2$. Such
a solution based on perturbation theory for weak disorder is
discussed below and applied to study the disorder-enhanced
tunnelling.  The relative merits of the two types of solutions for
describing the disorder-enhanced tunnelling are discussed towards
the end of the paper.

On the other hand, we close by pointing out that the
non-perturbative invariant imbedding solution for a tunnelling
barrier in \cite{5}, which does lead to the phenomenon of
disorder-enhanced tunnelling, also leads to the exponential growth
of resistance and conductance of a disordered barrier discussed
earlier by Freilikher et al.\cite{1}.

We consider an electron of energy $E=\hbar^2k^2/2m$ (with units such
that $\hbar=m=1$) which impinges from the right on a random
one-dimensional tunnel barrier

\begin{equation}\label{eq1}
V(x)=V+v(x) \quad ,
\end{equation}
confined to the region $0\leq x\leq L$. $V$ denotes the mean of
$V(x)$ and $v(x)$ is a weak gaussian white-noise,

\begin{equation}\label{eq2}
\langle v(x) v(x') \rangle =\xi\delta (x-x'), \langle v(x) \rangle
=0 \quad .
\end{equation}
Outside the barrier the particle is described by the wavefunction

\begin{subequations}\label{eq3}
\begin{align}
\psi (x)&= e^{-ik(x-L)}+r(L) e^{ik(x-L)}, x>L\quad ,
\label{eq3a}\\
\psi (x)&= t(L)e^{-ikx}, x<0 \quad ,\label{eq3b}
\end{align}
\end{subequations}
where the complex reflection and transmission coefficient amplitudes
$r(L)\equiv r$ and $t(L)\equiv t$ are determined by the invariant
imbedding equations\cite{7}

\begin{equation}\label{eq4}
ik\frac{dr(L)}{dL}= -2k^2 r(L) +V(L) (1+r(L))^2 \quad ,
\end{equation}

\begin{equation}\label{eq5}
ik\frac{dt(L)}{dL}= -k^2 t(L) +V(L) (1+r(L)) t(L) \quad .
\end{equation}
In I we discussed a useful approximate solution of \eqref{eq4} valid
within some energy interval around the value $E=V/2$ such that for
the most typical values of \eqref{eq1} one has

\begin{equation}\label{eq6}
2E-V\simeq v(L) \quad .
\end{equation}
In this case the r.h.s. of \eqref{eq4} is approximately $V(L)
(1+r(L)^2)$ and the solution of \eqref{eq4} subject to $r(0)=0$
is\cite{5}

\begin{equation}\label{eq7}
r(L)=-i \tanh [\frac{1}{k}\int^L_0 dL' V(L')] \quad .
\end{equation}
By inserting \eqref{eq7} in \eqref{eq5} we obtain the corresponding
exact solution for the amplitude transmission coefficient,

\begin{equation}\label{eq8}
t(L)= \frac{e^{i(kL-\int^L_0 dL' V(L'))}} {\cosh[\frac{1}{k}\int^L_0
dL' V(L')]} \quad ,
\end{equation}
which is valid for energies of the incident electron close to $V/2$.
Note that (\ref{eq7}-\ref{eq8}) verify probability conservation,

\begin{equation}\label{eq9}
\vert r(L)\vert^2+\vert t(L)\vert^2=1 \quad ,
\end{equation}
as required.

The solution \eqref{eq7} of the invariant imbedding equation
\eqref{eq4} for $E\simeq V/2$ has been further discussed by Haley
and Erd\"{o}s\cite{10} in the context of the resistance, $\rho (L)$,
of the disordered barrier defined by the Landauer formula

\begin{equation}\label{eq10}
\rho (L)= \frac{\vert r(L)\vert^2}{ 1-\vert r(L)\vert^2} \quad .
\end{equation}
These authors also developed a more general result for the Landauer
resistance of a disordered barrier valid for any value of $V/E$.

We first address the question of the disorder enhanced tunnelling
across the potential barrier \eqref{eq1}, for weak disorder.  More
precisely, by expanding $\vert t(L)\vert^2$ in \eqref{eq8} to second
order in $k^{-1}\int^L_0 d L' v(L')$ assumed to be small and
averaging over the disorder using \eqref{eq2}, we get

\begin{equation}\label{eq11}
\langle\vert t(L)\vert^2\rangle=\frac{1}{\cosh^2\frac{VL}{k}} \left[
1+\frac{\xi L}{k^2} \left(3\tanh^2 \frac{VL}{k}-1\right) \right]
\quad .
\end{equation}
It follows that the sign of the effect of the disorder on the
transmittance of the barrier is given by the sign of the factor
$3\tanh^2 (VL/k)-1$.  Thus, we find that for parameters such that
$\exp (\frac{-2VL}{k})<2-\sqrt 3$, the transmission coefficient of
the disordered barrier is enhanced by the effect of weak disorder.
The threshold value

\begin{equation}\label{eq12}
\frac{VL}{k}=\sqrt V L= kL=-\frac{1}{2}\ln (2-\sqrt 3)\simeq 0,659
\quad ,
\end{equation}
above which the mean transmission coefficient \eqref{eq11} increases
with increasing disorder is in reasonable agreement with the
critical value $kL\simeq 0,58$ at which the effect of the disorder
in the transmittance changes sign in the results of fig. 2 of
Ref.\cite{3}.  Below this threshold value of $kL$
 the disorder reduces the transmittance while above it the disorder enhances it.

On the other hand, following Kim et al., we define the effective
disorder parameter

\begin{equation}\label{eq19}
g=\frac{\xi}{k^3} \quad ,
\end{equation}
in terms of which we obtain from \eqref{eq11} (with $V=2E=k^2$)

\begin{equation}\label{eq20}
\frac{ d\langle\vert t(L)\vert\rangle^2}{dg} =\frac{kL}{\cosh^2 kL}
(3\tanh^2 kL-1) \quad .
\end{equation}
This defines the initial rate (i.e. near $g=0$) of variation of the
transmittance as a universal function of the scaled wavenumber $kL$,
both above and below the threshold.  The expression \eqref{eq20} may
be compared with the slopes at the origin of the transmittances as a
function of $g$ in fig. 2 of \cite{3}, for various values of $kL$.
This comparison is shown in table 1, indicating a quite reasonable
agreement between the two sets of results.

The enhancement of the transmittance of the tunnel barrier for weak
disorder suggests, of course, a non-monotonic variation at larger
disorder since beyond sufficiently large disorder the transmittance
necessarily decreases to zero.  This follows e.g. from the form of
the typical transmission coefficient obtained from \eqref{eq8}:

\begin{align}\label{eq13}
\vert t(L)\vert^2_{\text{typical}}&= \left(\langle \cosh^2 \left[
\frac{1}{k}\int^L_0 dL' V(L') \right]
\rangle\right)^{-1}\nonumber\\
&= 2 \left[ 1+e^{\frac{2\xi L}{k^2}} \cosh \frac{2VL}{k}
\right]^{-1} \quad ,
\end{align}
which vanishes for $\xi\rightarrow \infty$.  The result \eqref{eq13}
is obtained by using the well-known formula\cite{11}

\begin{equation}\label{eq14}
\langle\exp\left[\pm a\int^L_0 v(L') dL'\right]\rangle =\exp
\left(\frac{a^2\xi L}{2}\right) \quad ,
\end{equation}
for averages over Gaussian correlated variables defined by
\eqref{eq2}.  The non-monotonic behaviour of the transmittance as a
function of disorder is revealed in detail by the numerical
calculations in Refs \cite{3,4}.

We now turn to the discussion of a new simple approximate solution
of the invariant imbedding equations, which in contrast to
(\ref{eq7}-\ref{eq8}), will be valid for arbitrary $E<V$. We first
rewrite (\ref{eq4}-\ref{eq5}) in terms of new amplitudes

\begin{equation}\label{eq21}
q(L)= e^{-2ikL} r(L), s(L)= e^{-ikL} t(L),
\end{equation}
which leads to

\begin{equation}\label{eq22}
ik\frac{d q(L)}{dL}= V(L)\left(e^{-ikL}+ e^{ikL} q(L)\right)^2 \quad
,
\end{equation}

\begin{equation}\label{eq23}
ik\frac{d s(L)}{dL}= V(L)\left(1+ e^{2ikL} q(L)\right) s(L) \quad .
\end{equation}
For $kL << 1$ these equations reduce approximately to

\begin{equation}\label{eq24}
ik\frac{d q(L)}{dL}\simeq V(L)\left (1+ q(L)\right)^2 \quad ,
\end{equation}

\begin{equation}\label{eq25}
ik\frac{d s(L)}{dL}\simeq V(L) \left (1+ q(L)\right)s(L), kL <<1
\quad ,
\end{equation}
and may be readily solved with the boundary conditions $q(0)=0$ and
$s(0)=1$. We find

\begin{equation}\label{eq26}
q(L)= \frac{1}{ik} \left(\int^L_0 dL'V(L')\right)
\left[1-\frac{1}{ik}\int^L_0 dL'V(L')\right]^{-1} \quad ,
\end{equation}

\begin{equation}\label{eq27}
s(L)= \left[1-\frac{1}{ik}\int^L_0 dL'V(L')\right]^{-1}, kL<<1\quad
.
\end{equation}
Restricting our attention to the transmittance, $\vert
t(L)\vert^2=\vert s(L)\vert^2$ we obtain from \eqref{eq1},
\eqref{eq21} and \eqref{eq27}

\begin{equation}\label{eq28}
\vert t(L)\vert^2 \frac{1}{1+\frac{1}{k^2}\left(VL+\int^L_0
dL'V(L')\right)^2} \quad .
\end{equation}
By expanding \eqref{eq28} to second order in the weak disorder
$v(L')$ in \eqref{eq1} and averaging the resulting expression using
\eqref{eq2} we finally obtain

\begin{equation}\label{eq29}
\langle\vert t(L)\vert^2\rangle= \frac{1}{Q(L)} \left[1+\frac{\xi
L}{k^2 Q(L)} \left(\frac{4 V^2L^2}{k^2 Q(L)}-1\right)\right] \quad,
\end{equation}
with
\begin{equation}\label{eq30}
Q(L)= 1+\left(\frac{VL}{k}\right)^2\quad .
\end{equation}

The expressions (\ref{eq29}-\ref{eq30}) show the existence of a
wavenumber domain defined by $3\left(\frac{VL}{k}\right)^2>1$ i.e.

\begin{equation}\label{eq31}
kL>\frac{2}{\sqrt 3}\frac{E}{V} \quad ,
\end{equation}
where the transmittance is enhanced by the disorder.  The critical
scaled wavenumber threshold, $\frac{2}{\sqrt 3}\frac{E}{V}$, may be
readily compared with the critical thresholds for disorder enhanced
transmission obtained numerically by Kim et al.\cite{3} and
displayed in their figures 3,4,5, successively for $V/E=1,5$, $
V/E=2$ and $ V/E=3$.  From these figures we obtain the critical
values $kL\simeq 0.8$ ($V/E=1,5$), $kL\simeq 0.58$ ($ V/E=2$) and
$kL\simeq 0.4$ ($ V/E=3$), whose comparison with the results
$kL=0,77$ ($V/E=1.5$), $kL=0,577$ ($V/E=2$) and $kL=0.385$ ($V/E=3$)
obtained from \eqref{eq31} shows remarkable agreement.

Finally we discuss the initial slopes, $d\langle\vert
t(L)\vert^2\rangle/dg\vert_{g=0}$, of the mean transmittance
\eqref{eq29} as a function of the disorder parameter \eqref{eq19},
for various scaled incident wavenumbers $kL$.  For the case $V/E=2$
where numerical results for the transmittance are available  in fig.
2 of Kim et al.\cite{3}, we get

\begin{equation}\label{eq32}
\frac{d\langle\vert t(L)\vert^2\rangle}{dg}=
\frac{kL(3(kL)^2-1)}{(1+(kL)^2)^3}, kL<<1 \quad .
\end{equation}
As shown in table I, the agreement between the numerical results
obtained from \eqref{eq32} with those inferred from the results of
fig. 2 of Kim et al.\cite{3} for various $kL$ is again quite
reasonable.

In conclusion, we have presented two complementary mathematical
treatments demonstrating the existence of scaled wavenumber
thresholds for the appearance of disorder-enhanced tunnelling of an
electron and allowing furthermore to calculate analytically the
initial rate of variation of transmittance with the disorder, as a
function of the incident wavenumber.  Our results are in good
agreement with extensive numerical calculations of Kim et
al.\cite{3}.

We close with a brief comparison of our results for the special
energy \eqref{eq6} for large $L$ with the results of Freilikher et
al.\cite{1} indicating that a weak disorder increases both the mean
resistance and the mean conductance of the tunnel barrier. The
resistance $(\rho)$ of the barrier defined by the four-probe
Landauer formula is

\begin{align}\label{eq15}
\rho &=
\frac{\vert r(L) \vert^2}{\vert t(L) \vert^2},\nonumber\\
&=\sinh^2 \left[\frac{1}{k}\int^L_0 dL' V(L')\right] \quad ,
\end{align}
using (\ref{eq7}-\ref{eq9}).  For the conductance, $g$, the
following two-probe formula

\begin{equation}\label{eq16}
g=\vert t(L)\vert^2 \quad ,
\end{equation}
is to be preferred, following well-known arguments\cite{12}.

From \eqref{eq2}, \eqref{eq14} and \eqref{eq15} we obtain at long
lengths such that $VL/k>>1$ (with $T_0(L)=4\exp (-2 VL/k)$)

\begin{equation}\label{eq17}
\langle\rho\rangle\simeq \frac{1}{T_0(L)} e^{\frac{2\xi L}{k^2}}
\quad ,
\end{equation}
and from \eqref{eq8} and \eqref{eq16}

\begin{equation}\label{eq18}
\langle g\rangle= \langle \vert t(L)\vert^2\rangle\simeq T_0 (L)
e^{\frac{2\xi L}{k^2}} \quad .
\end{equation}
These expressions are similar to the results obtained earlier by
Freilikher et al.\cite{1}, using a different method.

As a final remark, we highlight the advantage of the approximate
analytic solution (\ref{eq7}-\ref{eq8}) of the invariant imbedding
equations at the energy of half the tunnelling barrier height, in
the context of the present work.  On the one hand, this solution has
enabled us to study the disorder enhanced transmittance at
wavenumbers larger than those which are accessible by means of the
perturbation analysis discussed above (see in particular table 1),
which requires $kL<<1$.  On the other hand, its non-perturbative
character is crucial for demonstrating the exponential growth of
both the resistance and the conductance shown previously in
\cite{1}.

\begin{table}[htbp]
\begin{center}
\begin{tabular}{|c|c|c|c|}
\hline
$kL$ & Kim et al.\cite{3} & Eq. (\ref{eq20}) & Eq. (\ref{eq32})\\
\hline
0,3 & -0,194 & -0,205 & -0,169\\
0,577 (0,58) & 0 & -0,076 & 0\\
0,6 & 0,025 & -0,057 & 0,019\\
0,659 & - & 0 & 0,068\\
1 & 0,414 & 0,311 & 0,25\\
1,5 & 0,557 & 0,395 & - \\
3 & 0,206 & 0,058 & -\\
\hline
\end{tabular}
\end{center}
\caption{Initial ($g=0$) slopes of transmittances as a function of
disorder parameter $g$, for various $kL$}
\end{table}
\end{document}